\documentclass[preprint]{aastex}

\pdfoutput=0
\usepackage{graphicx}
\usepackage{dcolumn}
\usepackage{amssymb} 
\usepackage{ctable}
\usepackage{bm}
\usepackage{rotating}
\usepackage{natbib}
\usepackage{array}

\newcommand{\cmnt}[1]{}

\def\fnl{f_{\rm NL}}
\def\vecp{\mathbf{p}}
\def\estp{\mathbf{\hat{p}}}
\def\matf{\mathbf{F}}
\def\vecq{\mathbf{q}}
\def\vecx{\mathbf{x}}
\def\matc{\mathbf{C}}
\def\mats{\mathbf{S}}
\def\matn{\mathbf{N}}

\def\hatn{\mathbf{\hat n}}

\def\VEV#1{\left\langle #1 \right\rangle}

\begin{document}
\bibliographystyle{plainnat}

\title{Systematic effects in large-scale angular power spectra of photometric quasars and implications for constraining primordial nongaussianity}



\author{Anthony R. Pullen}
\affil{NASA Jet Propulsion Laboratory, California Institute of Technology, 4800 Oak Grove Drive, MS 169-234, Pasadena, California, U.S.A.}
\email{anthony.r.pullen@jpl.nasa.gov}
\and
\author{Christopher M. Hirata}
\affil{Department of Physics, California Institute of Technology, Mail Code 350-17, Pasadena, CA 91125, USA}

\begin{abstract}
Primordial non-Gaussianity of local type is predicted to lead to enhanced halo clustering on very large scales.  Photometric quasars, which can be seen from cosmological redshifts $z>2$ even in wide-shallow optical surveys, are promising tracers for constraining non-Gaussianity using this effect.  However, large-scale systematics can also mimic this signature of non-Gaussianity.  In order to assess the contribution of systematic effects, we cross-correlate overdensity maps of photometric quasars from the {\it Sloan Digital Sky Survey} (SDSS) Data Release 6 (DR6) in different redshift ranges.  We find that the maps are significantly correlated on large scales, even though we expect the angular distributions of quasars at different redshifts to be uncorrelated. This implies that the quasar maps are contaminated with systematic errors.  We investigate the use of external templates that provide information on the spatial dependence of potential systematic errors to reduce the level of spurious clustering in the quasar data. We find that templates associated with stellar density, the stellar color locus, airmass, and seeing are major contaminants of the quasar maps, with seeing having the largest effect.  Using template projection, we are able to decrease the significance of the cross-correlation measurement on the largest scales from 9.2$\sigma$ to 5.4$\sigma$.  Although this is an improvement, the remaining cross-correlation suggests the contamination in this quasar sample is too great to allow a competitive constraint on $f_{\rm NL}$ by correlations internal to this sample.  The SDSS quasar catalog exhibits spurious number density fluctuations of $\sim$2\% rms, and we need a contamination level less than 1\% (0.6\%) in order to measure values of $\fnl$ less than 100 (10).  Properly dealing with these systematics will be paramount for future large scale structure surveys that seek to constrain non-Gaussianity.
\end{abstract}

\section{Introduction} \label{S:intro}

Inflation is the standard paradigm for the generation of perturbations in matter density that produces large-scale structure~(LSS) in the Universe \citep{infl1,Guth:1980zm,Linde:1981mu,Albrecht:1982wi}.  Although the inflationary paradigm has successfully predicted various properties of the observable universe, including flatness and a nearly scale-invariant spectrum of perturbations \citep{Mukhanov:1981xt,Hawking:1982cz,Guth:1982ec,infl2,Bardeen:1983qw}, the correct model of inflation has yet to be confirmed.  The simplest inflation models predict nearly Gaussian primordial perturbations, though more complex models such as multi-field inflation \citep{Linde:1996gt,Bernardeau:2002jy,Lyth:2002my} posit a departure from a Gaussian distribution.  Alternatives to inflation, such as the ekpyrotic model of cyclic expansion and contraction \citep{Khoury:2001wf,Steinhardt:2002ih,Creminelli:2007aq}, also predict non-Gaussian primordial perturbations.  Since a detection of non-Gaussianity would discriminate between these fundamentally different models, much work is being done to constrain non-Gaussianity, both in LSS through the galaxy distribution and through anisotropies in the cosmic microwave background (CMB).  Primordial non-Gaussianity is readily probed through measurements of the bispectrum of the CMB, in which a nonzero measurement constitutes a ``smoking gun'' detection, modulo any systematic effects.  Some alternative probes of non-Gaussianity in LSS include the galaxy bispectrum, which is plagued by nonlinearities,  and galaxy cluster abundances and dark-matter halo clustering, which suffer from low-number statistics.  The accepted parametrization of primordial non-Gaussianity is to introduce a quadratic term to the primordial potential $\Phi$, written as
\begin{eqnarray}\label{E:fnl1}
\Phi = \phi+f_{\rm NL}(\phi^2-\VEV{\phi^2})\, ,
\end{eqnarray}
where $\phi$ is a Gaussian random field \citep{Komatsu:2001rj,Gangui:1993tt}.  This form describes local-type non-Gaussianity with an amplitude $f_{\rm NL}$.  The latest constraint on $f_{\rm NL}$ is from the Wilkinson Microwave Anisotropy Probe's (WMAP) Seven-Year bispectrum, which gives $-10<f_{\rm NL}<+74$ at 95\% C.L.~\citep{Komatsu:2010fb}.  \emph{Planck} \citep{planck}, which will soon give results from its first data release, is expected to produce constraints on $f_{\rm NL}$ of order $\sigma(f_{\rm NL})\sim 7$ \citep{Cooray:2008xz}.

One useful effect of non-Gaussianity that has gained much attention, and which we explore in this analysis, is a distinct scale-dependent clustering bias on large scales \citep{Dalal:2007cu,Slosar:2008hx}.  Specifically, it has been shown that $f_{\rm NL}$-type non-Gaussianity produces a shift in the bias that behaves as $\Delta b(k)\propto f_{\rm NL}/k^2$; this can suppress or enhance clustering on the largest scales. Various authors have used the large-scale angular power spectra of LSS tracers to constrain $f_{\rm NL}$ assuming a scale-dependent bias.  \citet{Slosar:2008hx}, using the Sloan Digital Sky Survey (SDSS) \citep{York:2000gk} Data Release 5 (DR5) \citep{AdelmanMcCarthy:2007wh}, derived the constraints $-82<f_{\rm NL}<+70$ at 95\% C.L.~using the photometric quasar sample and  $-29<f_{\rm NL}<+70$ at 95\% C.L.~when these quasars are combined with other data sets, such as the integrated Sachs-Wolfe effect (ISW) and luminous red galaxies (LRGs).  This analysis was extended in \citet{Xia:2011} to include the SDSS DR6 \citep{AdelmanMcCarthy:2007wu} photometric quasar catalog compiled by~\citet{Richards:2008eq}, hereafter RQCat.  \citet{Tseliakhovich:2010kf} extended this analysis to a two-parameter curvaton model \citep{Linde:1996gt,Mollerach:1989hu,Lyth:2001nq,Lyth:2002my,Boubekeur:2005fj,Huang:2008zj} while others have sought to use this method in combination with other data sets \citep{Xia:2010yu,DeBernardis:2010kc}.  \citet{DeBernardis:2010kc} also showed that \emph{Planck} and {\it Euclid} \citep{Amiaux:2012} together could possibly detect $f_{\rm NL}\sim 5$.  Finally, work on using multiple tracers of different bias \citep{Hamaus:2011} has predicted future limits of $\fnl\sim 0.1$.

As in all auto-spectrum measurements, particularly on large angular scales, potential correlations caused by systematic effects must be removed to isolate the correlation signal from the LSS tracer.  Systematics have been a concern in recent analyses of the {\it Baryon Oscillation Spectroscopic Survey} (BOSS) \citep{Eisenstein:2011}.  Studies of BOSS photometric galaxies \citep{Ross:2011, Ho:2012} attempted to remove systematic effects by fitting an amplitude for each systematic template, and subtracting its estimated contribution from the power spectrum.  Recent work in \citet{Huterer:2012} outlines a formalism for how photometric calibration variations affect power spectrum and cosmological parameter estimations.  Another method for removing systematics is mode projection, which explicitly marginalizes over the amplitude of each template that traces the spatial dependence of a potential systematic error. This method has been profitably used for foreground removal in the low-$\ell$ CMB power spectrum (e.g. \citealt{Slosar:2004fr}). This method is more robust than subtracting the estimated contamination from the power spectrum estimator, because it is {\em exactly} insensitive to any systematic that traces the spatial dependence of the template. It also avoids the possibility of ``oversubtraction'' -- the phenomenon in which chance correlations of the real galaxy density with systematics templates lead to an overestimate of the contamination, and hence lead to a downward bias in the low-$\ell$ power spectrum.

In this paper, we attempt to remove large-scale systematics from the latest photometric quasar sample available, which is RQCat. We investigate the photometric quasar sample because, by probing large redshifts, quasars are able to probe scale-dependent bias more effectively than other matter tracers, as demonstrated in \citet{Slosar:2008hx}.  We use photometric quasars from a larger redshift range than this previous analysis, using redshifts as low as $z=0.9$ and as high as $z=2.9$.  We divide this sample into three redshift slices (0.9--1.3, 1.6--2.0, 2.3--2.9) and construct three angular cross-power spectra to test if they are consistent with zero, which we expect since quasars at different redshifts should be uncorrelated on large scales.  We also construct templates of various potential systematic effects and try mode-projecting them from the angular cross-power spectrum estimator.

Note that the sample used by e.g. \citet{Slosar:2008hx} was constructed by a different algorithm and is {\em not} the restriction of the RQCat sample to an earlier, smaller footprint. Therefore the specific amplitude of systematic errors herein should not be taken as quantitatively indicative of the level of contamination in these previous analyses.

We find that without mode-projecting any systematics, the first and second redshift slices are correlated with an 9.2$\sigma$ significance, and the first and third redshift slices are correlated with a 2.7$\sigma$ significance, both on large scales.  This implies that systematics are heavily contaminating large-scale correlations in the quasar maps.  Using mode projection, we find significant reductions in the correlations between redshift slices after removing templates corresponding to the stellar density, stellar locus offsets, airmass, and seeing.
We are able to decrease the significance of the correlation between the first and third redshift slices to 1.8$\sigma$; however, the significance of the correlation between the first and second redshift slices decreased only to 5.4$\sigma$, which is indicative of significant contamination.  While the quasar catalog exhibits $\sim$2\% contamination, measuring $\fnl$ will require a contamination level less than 1\%.  We conclude that other systematic effects are still present that are contaminating the SDSS photometric quasar sample.  We must improve methods of construct cleaner photometric quasar samples if we hope to use them to provide competitive constraints on $\fnl$ on large scales.

The plan of our paper is as follows: in Sec.~\ref{S:data} we describe the photometric quasar sample we use in the analysis.  In Sec.~\ref{S:theory} we describe briefly the theory behind scale-dependent bias and how it affects the angular power spectrum.  In Sec.~\ref{S:method}, we outline the estimators we use, the method of mode projection, and the systematics for which we search.  In Sec.~\ref{S:results}, we give the angular cross-power spectrum measurements, before and after mode projection, and we discuss its implications.  We conclude in Sec.~\ref{S:conclude}.  Wherever not explicitly mentioned, we assume for the background cosmology and power spectrum a flat $\Lambda$CDM cosmology with parameters compatible with the WMAP7 data release \citep{Larson:2010gs}.

\section{Choice of Sample} \label{S:data}

We use the photometric quasars from the SDSS DR6 \citep{AdelmanMcCarthy:2007wu} to trace the matter density and construct its angular spectrum.  The SDSS consists of a 2.5-m telescope \citep{Gunn:2006tw} with a 5-filter (\textit{ugriz}) imaging camera \citep{Gunn:1998vh} and a spectrograph.  Automated pipelines are responsible for the astrometric solution \citep{Pier:2002iq} and photometric calibration \citep{Fukugita:1996qt, Hogg:2001gc, Tucker:2006dv, Padmanabhan:2007zd}.  Bright galaxies, luminous red galaxies (LRGs), and quasars are selected for follow-up spectroscopy \citep{Strauss:2002dj, Eisenstein:2001cq, Richards:2002bb, Blanton:2001yk}.  The data used here were acquired between August 1998 and June 2006 and are included in SDSS Data Release 6 \citep{AdelmanMcCarthy:2007wu}.

Specifically, we use the photometric quasar catalog (RQCat) composed by \citet{Richards:2008eq}.  The entire catalog consists of 1,172,157 objects from 8417 deg$^2$ on the sky selected as quasars from the SDSS DR6 photometric imaging data.  Quasars are the brightest objects at large redshifts ($z>1$), making them better tracers of the matter density at large scales than LRGs.  We limit our dataset in this analysis to UV-excess ($u-g<1.0$) quasars.  Specifically, we implement this choice by requiring the catalog columns \textbf{good} $> 0$ and \textbf{uvxts} = 1.  We divide quasars into 3 redshift slices $(z_{\rm p,min},z_{\rm p,max})=(0.9,1.3),\,(1.6,2.0)$, and $(2.3,2.9)$.  We plot the three redshift distributions in Fig.~\ref{F:redplot1}, while their properties are given in Table \ref{T:redshifts1}.  The procedure for constructing the redshift distributions is described in the appendix and is similar to that described in Ref.~\citep{Ho:2008bz}.  For the survey geometry we use the DR6 survey mask as a union of the survey runs retrieved from the SDSS CAS server.  We omitted runs 2189 and 2190 because many objects in these runs were cut from RQCat.  This mask was pixelized using the MANGLE software \citep{Hamilton:2003ea,Swanson:2007aj}.  We pixelize the quasars as a number overdensity, $\delta_q=(n-\overline{n})/\overline{n}$, onto a HEALPix pixelization \citep{Gorski:2004by} of the sphere with $N_{res}=256$, where $n$ is the pixel's number of quasars divided by the pixel's survey coverage $w$ and $\overline{n}=(\sum_in_iw_i)/(\sum_iw_i)$.

\begin{figure}
\begin{center}
\includegraphics[width=5.5in]{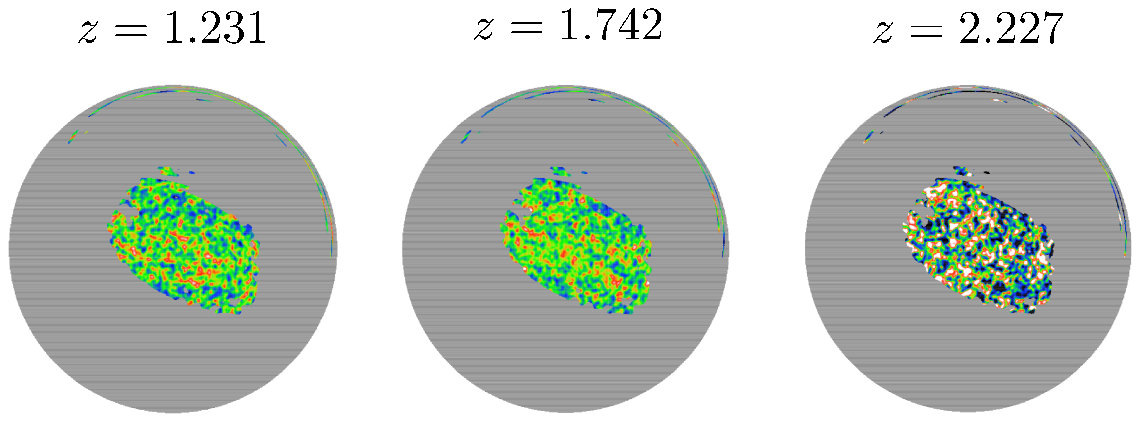}
\includegraphics[width=1.6in]{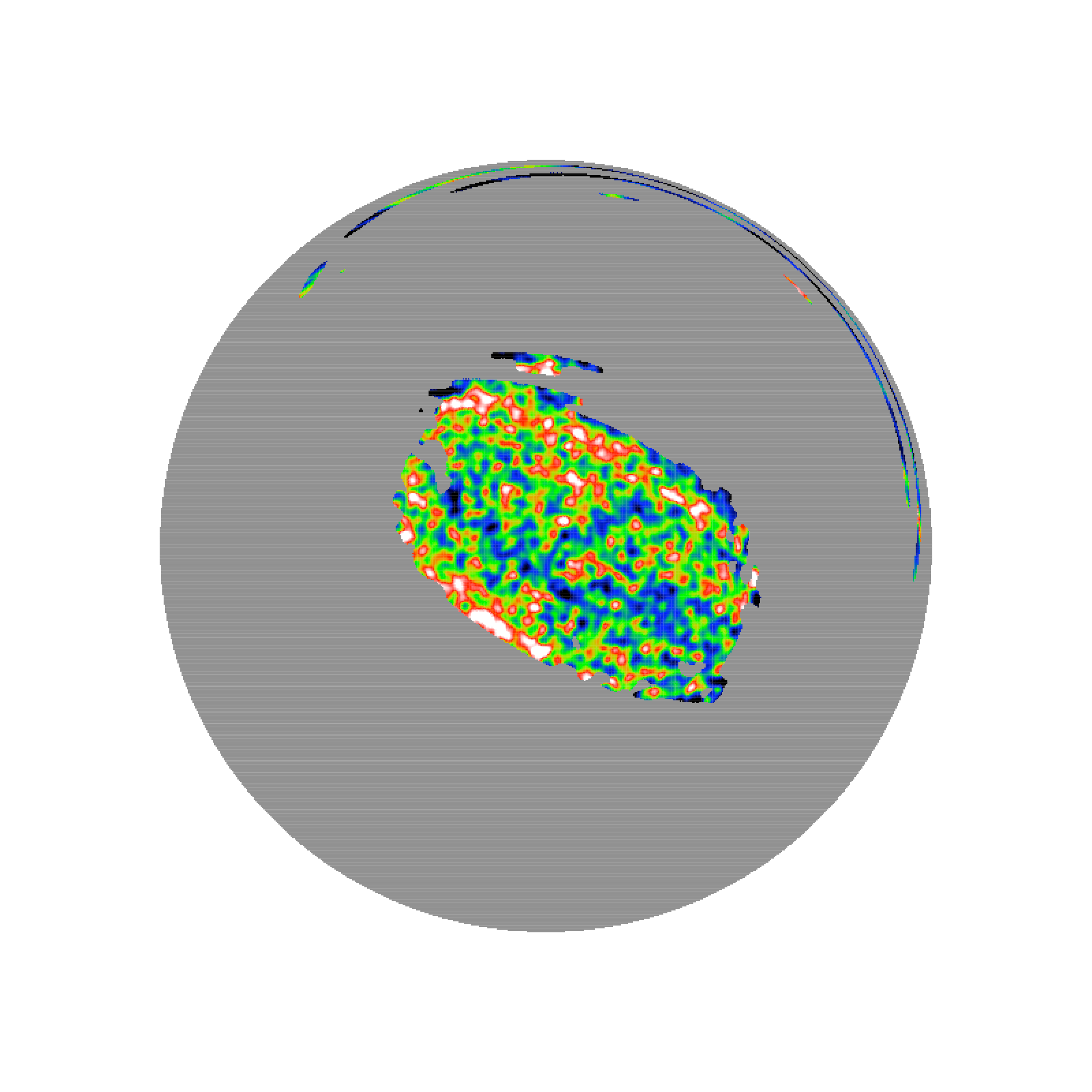}
\caption[The QSO density in the 3 photometric redshift slices.]{\label{F:qsomaps}{\it Top}: The QSO density in the photometric redshift slices z01, z02, and z03, respectively.  The 180$^\circ$ radius caps are displayed in a Lambert Azimuthal Equal-Area Projection, with the North Galactic Pole at the map center, $l=0^\circ$ at right, and $l=90^\circ$ at bottom.  {\it Bottom}: The QSO density in redshift slice z03 with no KDE QSO density cut.  Notice the large-scales fluctuations that would ruin scale-dependent bias measurements.}
\end{center}
\end{figure}

We try to clean our sample by removing pixels that we suspect are contaminated.  We reject pixels with extinction $E(B-V)\geq 0.05$, full widths at half-maximum of its point-spread function (PSF) FWHM $\geq 2$ arcsec, and stellar densities (smoothed with a $2^\circ$ FWHM Gaussian profile) $n_{stars}\geq 562$ stars/deg$^2$ (twice the average stellar density), similar to \citet{Ho:2008bz}.  The extinction cut is very important because a high extinction affects the $u$ band, which is crucial to identifying quasars.  Also, since stars tend to be misidentified as quasars, it seems prudent to cut regions with high stellar density.  We implement these cuts using dust maps from \citet{Schlegel:1997yv} and a stellar overdensity map, smoothed with a 2$^\circ$ FWHM Gaussian profile, constructed from stars ($18.0<r<18.5$) with $i<21.3$ (cutoff for point sources in RQCat) from the SDSS DR6 \citep{AdelmanMcCarthy:2007wu}.  We also reject pixels for which the survey region covers less than 80\% of the pixel area. We also excised several rectangular regions that appeared to contain missing data; the angular rectangles in equatorial (J2000) coordinates that were removed are $(\alpha,\delta)=(122^\circ\mbox{--}139^\circ,-1.5^\circ\mbox{--}(-0.5)^\circ)$, $(121^\circ\mbox{--}126^\circ,0^\circ\mbox{--}4^\circ)$, $(119^\circ\mbox{--}128^\circ,4^\circ\mbox{--}6^\circ)$, $(111^\circ\mbox{--}119^\circ,6^\circ\mbox{--}25^\circ)$, $(111.5^\circ\mbox{--}117.5^\circ,25^\circ\mbox{--}30^\circ)$, $(110^\circ\mbox{--}116^\circ,32^\circ\mbox{--}35^\circ)$, $(246^\circ\mbox{--}251^\circ,8.5^\circ\mbox{--}13.5^\circ)$, $(255^\circ\mbox{--}270^\circ,20^\circ\mbox{--}40^\circ)$,\\ $(268^\circ\mbox{--}271^\circ,46^\circ\mbox{--}49^\circ)$, and $(232^\circ\mbox{--}240^\circ,26^\circ\mbox{--}30^\circ)$.  Finally, we cut regions comprising HEALPix $N_{\rm side}=64$ pixels that contained no stars for which to estimate the stellar color locus and flux-error locus offset systematic templates, which use photometric measurements of stars, as well as cut pixels with color locus offsets greater than 5 mag (see Sec.~\ref{S:system}).  After these cuts, the survey region comprises 122,780 pixels covering a solid angle of $6440$ deg$^2$.  The quasar maps for each slice are shown in Fig.~\ref{F:qsomaps}.  Note that we reduce a large-scale systematic by implementing an additional cut on the quasars in the highest redshift bin to exhibit a high kernel-density-estimator (KDE) quasar probability density, requiring the catalog column \textbf{qsodens} $> 0$ (see Fig.~\ref{F:qsomaps}).

\section{Theory} \label{S:theory}
\subsection{Scale-dependent bias due to a non-Gaussian primordial inhomogeneity}

The derivation for the scale-dependent bias model for non-Gaussianity of the ``local'' ($\fnl$) type is detailed in \citet{Slosar:2008hx}, for which we give a brief overview.  The model assumes the halo model scenario (see \citet{halorev} for a detailed review) in which all matter is contained in a distribution of halos on large scales which mimics the distribution of matter on small scales.  In the Gaussian case, fluctuations on small and large scales are uncorrelated.  However, a signature of local-type non-Gaussianity is that the $f_{\rm NL}\phi^2$ term in the gravitational potential (see Eq.~\ref{E:fnl1}) causes small-scale matter fluctuations to correlate with large-scale halo fluctuations due to mixing of their respective gravity perturbations.  Positive (negative) $f_{\rm NL}$ incurs a positive (negative) correlation between scales and an increase (decrease) in the halo bias on large scales as compared to the Gaussian case.  Specifically, this scale-dependent shift in the bias was derived for a general halo mass function $n(M)$ in \citet{Slosar:2008hx} to be
\begin{eqnarray}\label{E:dbgen}
\Delta b(M,k)=\frac{3\Omega_m H_0^2}{c^2 k^2 T(k) D(z)}f_{\rm NL}\frac{\partial\ln n}{\partial\ln\sigma_8}\, ,
\end{eqnarray}
where $k$ is the wavenumber corresponding to the length scale probed, $T(k)$ is the transfer function, $D(z)$ is the the growth factor normalized such that $D(z=0)=1$, $c$ is the speed of light, $H_0$ is the Hubble parameter today, and $\sigma_8$ is the rms overdensity in a sphere of radius $R=8h^{-1}$ Mpc.  For the case of a universal mass function, such as the Press-Schechter \citep{Press:1973iz} or Sheth-Tormen \citep{Sheth:1999mn} mass functions, this expression was shown in \citet{Slosar:2008hx} to reduce to
\begin{eqnarray}\label{E:dbuni}
\Delta b(M,k)=3f_{\rm NL}(b-p)\delta_c\frac{\Omega_m}{k^2 T(k) D(z)}\left(\frac{H_0}{c}\right)^2\, ,
\end{eqnarray}
the expression first derived in \citet{Dalal:2007cu}, where $\Omega_m$ is the matter density today relative to the critical density for a flat universe, and $\delta_c$ is the critical density of spherical collapse.  The parameter $p$ ranges from 1 for LRGs, which populate all halos equally, to 1.6 for quasars that populate only recently merged halos.  We use Eq.~(\ref{E:dbuni}) with $p=1.6$ in our analysis to model the scale-dependent bias of quasars.

\subsection{Angular power spectrum due to scale-dependent bias and non-Gaussianity} \label{S:cl}

\begin{figure}
\begin{center}
{\includegraphics[width=0.8\textwidth]{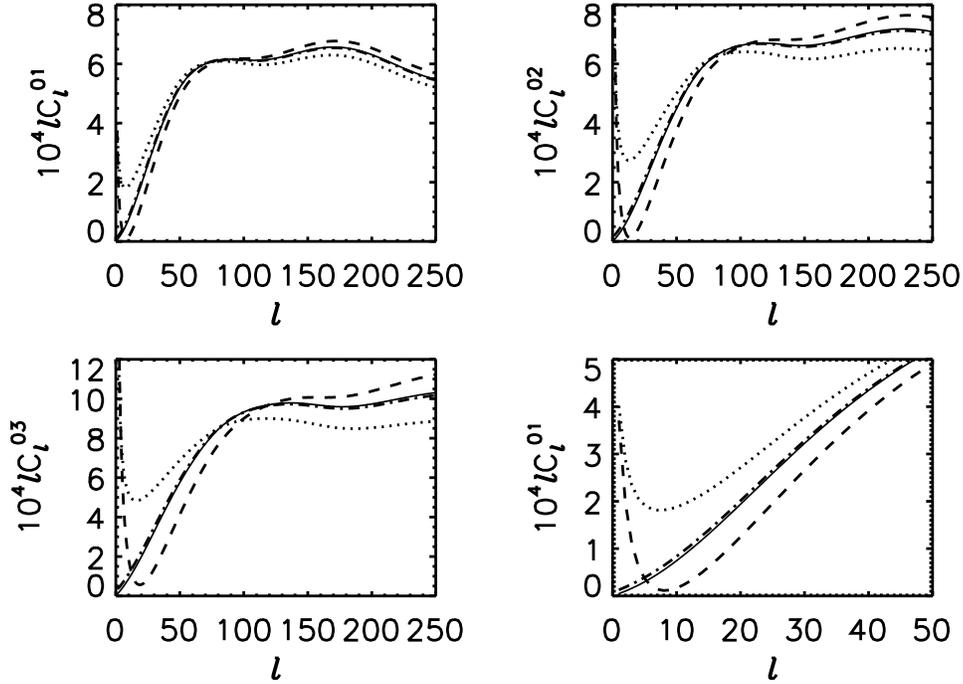}}
\caption[The predicted QSO angular power spectra for null and nonzero $f_{NL}$]{\label{F:cl} The predicted QSO angular power spectra for null and nonzero $f_{NL}$.  The top-left and bottom-right panels are both for the z01 redshift bin, except that the bottom-right panel's $\ell$-range ends at $\ell=50$ to see the low-$\ell$ behavior of the spectrum.  The solid lines are the spectra for the Gaussian case, and the dotted (dashed) lines represent the $\fnl=100\,(-100)$ case.  The dot-dashed line represents $\fnl=10$.  We normalized the $\fnl\neq0$ cases such that the variances of number density fluctuations within a pixel ($\propto\sum_\ell (2\ell+1)C_\ell$) for these cases are equal to the value for $\fnl=0$.}
\end{center}
\end{figure}

The clustering bias is a key component of the matter power spectrum $\Delta(k)$, thus we expect scale-dependent bias to affect it as well as the angular power spectrum $C_\ell$.  Since we are interested in large correlations in this analysis, we cannot use the small-scale Limber approximation, but must use the full expression for $C_\ell$ including redshift-space distortions. These expressions are given for the scale-independent bias case in Eqs.~29 and 30 in \citet{Padmanabhan:2006ku} and for the scale-dependent bias case in \citet{Slosar:2008hx}.  We assume the linear matter power spectrum in our analysis\footnote{\citet{Slosar:2008hx} confirmed that nonlinearities are negligible for $k<0.1h$ Mpc$^{-1}$.  For our redshifts, this corresponds to $\ell<270$.  We only use $C_\ell$ for $\ell<250$, so it is safe to use the linear matter power spectrum in our analysis.}.  The $C_\ell$s were calculated for all 3 redshift slices and presented in Fig.~\ref{F:cl}, including for the cases of nonzero $\fnl$.

\section{Methodology} \label{S:method}

Non-Gaussianity, which produces a shift in the bias of the form $\Delta b(k)\propto f_{\rm NL}/k^2$, manifests itself on large angular scales.  If any systematic effect can increase power on large scales, we will be unable to distinguish it from scale-dependent bias.  One way to attempt to assess the contamination of our quasar maps would be to cross-correlate the quasar maps with each of the systematics and see if it is nonzero, which was done in \citet{Slosar:2008hx} with stars and red stars.  However, this will be difficult to do for the other systematics we consider because the errors are uncertain.  Thus, in this analysis we compare angular cross-power spectra before and after mode-projecting an individual systematic to see if it decreases.  More specifically, we first calculate the angular cross-power spectrum without attempting to remove any systematics.  We then recalculate it, except that we mode-project one systematic.  Finally, we compare the two to see if the spectra are significantly different. Since there are three redshift bins, there will be 3 cross-spectra.

\begin{figure}
\begin{center}
\includegraphics[width=0.8\textwidth]{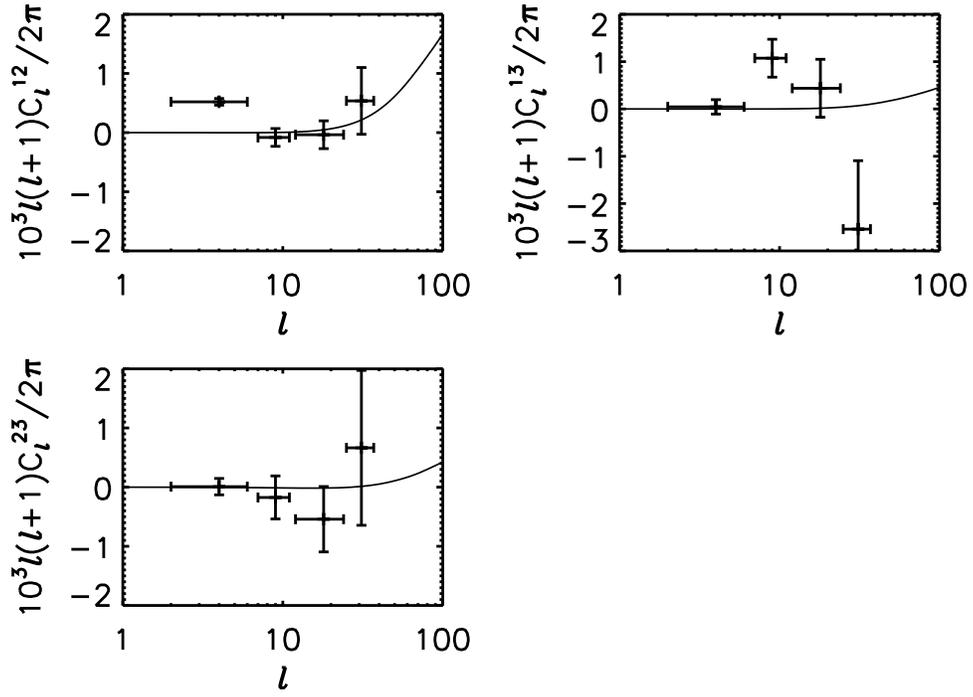}
\caption[The measured cross-correlation angular power spectra between all three QSO redshift slices]{\label{F:qcross} The measured cross-correlation angular power spectra between all three QSO redshift slices.  The crosses are the measured spectra with 1$\sigma$ errors and the solid lines are the predicted spectra.  Note that the predicted spectra are not exactly zero at smaller scales because the redshift distributions have small overlaps.}
\end{center}
\end{figure}

\begin{figure}
\begin{center}
\includegraphics[width=0.8\textwidth]{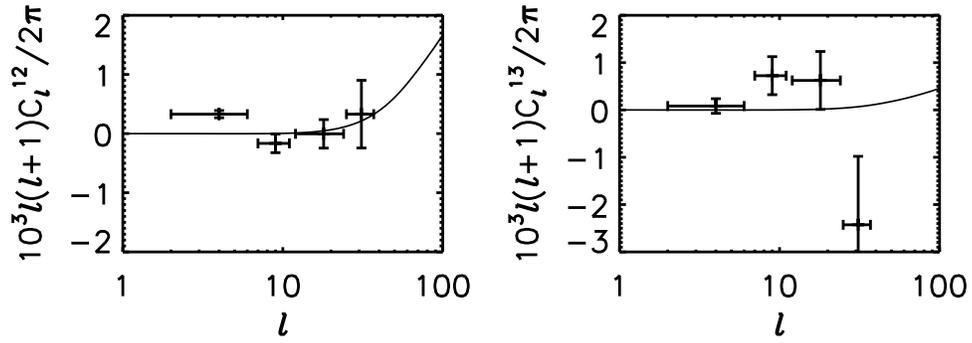}
\caption{\label{F:clqcrsys} The measured cross-correlation angular power spectra between z01 and z02 as well as z01 and z03 with systematic templates included.  The crosses are the measured spectra with 1$\sigma$ errors and the solid lines are the predicted spectra.  Note that the predicted spectra are not exactly zero at smaller scales because the redshift distributions have small overlaps.}
\end{center}
\end{figure}

The cross-spectra are useful because we expect different quasar maps to be uncorrelated if they are clean since objects from separate redshift bins should not correlate with each other.\footnote{Even with a null contamination, there will be a finite cross-correlation between the maps because the redshift distributions for the bins do overlap, but it is much smaller than the typical errors on the cross-correlation measurement.  We plot the fiducial cross-power spectra along with our cross-spectra estimates in Figs.~\ref{F:qcross} and \ref{F:clqcrsys}, where we see a finite cross-correlation on smaller scales.} However, if the same systematic effect appears in two quasar maps, then their cross-spectra will be non-zero.  So our goal for each cross-spectra is to identify those systematics for which mode-projection decreases the magnitude of the cross-spectra, and then try mode-projecting this set of systematics together to see if the cross-spectra goes to zero, which would be the limit of the usefulness of this test.  If it remains non-zero, then we know that there remains systematic effects in the maps that are unidentified.  Of course, this will only identify those systematics which contaminate two or all three quasar maps.  There may be systematic effects that contaminate only one of the maps that would not appear in the cross-spectra yet still contaminate any estimates of $\fnl$ using auto-spectra.  We decide not to search for this case since the auto-spectra we would need for the test would also be used to to constrain $\fnl$.  However, this case should be considered in future work.


\subsection{Angular power spectrum estimators}
In order to construct our angular power spectrum estimator, we begin by constructing the data vector
\begin{eqnarray}
\vecx=\left(\begin{array}{c}\vecx_1\\\vecx_2\end{array}\right)\, ,
\end{eqnarray}
where $\vecx_1$ and $\vecx_2$ length-$N_{\rm pix}$ vectors constituting the two maps we are cross-correlating, making $\vecx$ have a length of $2N_{\rm pix}$.  The covariance matrix for $\vecx$ is given by
\begin{eqnarray}
\matc=\left(\begin{array}{cc}\mats_1+\matn_1&\mathbf{0}\\\mathbf{0}&\mats_2+\matn_2\end{array}\right)\, ,
\end{eqnarray}
where $\mats_1$ and $\mats_2$ ($\matn_1$ and $\matn_2$) are the signal (noise) covariance matrices for Map 1 and Map 2, respectively, and we neglect any cross-correlation between the maps.  $\mats$ is given by
\begin{eqnarray}\label{E:clcovar}
S_{ij}=\sum_{\ell}\left(\frac{2\ell+1}{4\pi}\right)W_\ell C_\ell P_\ell[\cos(\hatn_i\cdot\hatn_j)]\, ,
\end{eqnarray}
where $C_\ell$ is the theoretical angular power spectrum (we assume $\fnl=0$ for our prior), $P_\ell(x)$ is a Legendre polynomial and $W_\ell$ is the pixel window function, and we assume Poisson noise for $\matn$ given by
$N_{ij} = \delta_{ij}/\overline{n}$
in terms of the mean number of galaxies per pixel $\overline{n}$.

We use a quadratic estimator \citep{Tegmark:1996qt, Padmanabhan:2002yv} to measure the angular power spectra in $N_{\rm bin}=4$ $\ell$-bins with ranges (2--6, 7--11, 12-24, 25--36).  Flat bandpowers allow us to write a parameter vector $\vecp=(\tilde{C}_1,\tilde{C}_2,\tilde{C}_3,\tilde{C}_4)$ where the band amplitudes $\tilde{C}_n$ are expressed as $C_\ell=\sum_{n=1}^{N_{\rm bin}}\tilde{C}_n\eta_\ell^n$, where $\eta_\ell^n$ is a step function that is 1 when $\ell$ is in bin $n$ and 0 otherwise.  Our estimator for $\vecp$ is constructed in the form
$\estp=\matf^{-1}\vecq$,
where
\begin{equation}\label{E:fq}
F_{ij} = \frac{1}{2}{\rm tr}\left[\matc,_i\matc^{-1}\matc,_j\matc^{-1}\right]
{\rm ~~~and~~~}
q_i = \frac{1}{2}\vecx^T\matc^{-1}\matc,_i\matc^{-1}\vecx
\end{equation}
are the Fisher matrix and quadratic estimator vector, respectively, and $\matc,_i = \partial\matc/\partial p_i$.  Note the matrix inversion and trace estimation are done by the iterative and stochastic methods described in detail in \citet{Hirata:2004rp} and \citet{Padmanabhan:2006ku}.\footnote{Such methods were introduced to cosmology by \citet{Oh:1998sr}.}

Note that we project out the monopole and dipole of the power spectra since these moments are not of interest cosmologically. The simplest way to remove a data point is to set its error to infinity (or a large value, so 1/error$^2=0$).  Since the errors for multipoles with low shot noise are proportional to the multipole, setting the monopole and dipole to large values actually removes their estimator contributions to $C_\ell/(C_\ell)^2 \sim 0$.  Giving large values to these moments increases their covariances in the estimator, making their contributions negligible, while increasing estimate errors slightly.  This method is a specific case of the more general projection method we use throughout the paper (see Sec.~\ref{S:modeproj}). This method was first presented in \citet{Rybicki:1992}, and was also presented in \citet{Tegmark:1996qt} and was applied to CMB foreground removal by \citet{Slosar:2004fr}.  Note this is not equivalent to constraining the monopole and dipole moments to large values.  This is performed in our analysis by setting the $C_\ell$ prior equal to 100 for $\ell=0$ and 1.

Our treatment of masked pixels is similar.  We construct our estimator in real space, such that masking pixels, or striking missing pixels from the covariance matrix before inversion, is equivalent to setting the noise in these pixels to infinity, which removes their effect from the estimator anyway.  It is true that masking causes the $C_\ell$ measurements to be correlated, but these correlations are described by the Fisher matrix $\matf$, such that $\matf^{-1}$ in the estimator expressed in Eq.~\ref{E:fq} removes the leakage, as shown in Sec.~3E of \citet{Tegmark:1996qt}.

\subsection{Mode projection} \label{S:modeproj}

We use the method of mode projection to remove modes due to systematics from our angular power spectrum estimators, instead of subtracting them directly, which could lead to spurious modes remaining.  This method projects out systematics by in effect giving these modes large noise values so that the estimator is rendered insensitive to them.  Mode projection \citep{Tegmark:1997yq,Bond:1998zw,Halverson:2001yy} has been used often for CMB data (see \citet{Halverson:2001yy} for a review), yet this method can also clean LSS analyses.  We begin with the assumption that the systematic's effect on the QSO map $\vecx$ is of the form
\begin{eqnarray} \label{E:sysmap}
\vecx^{\rm obs}=\vecx^{\rm true}+\sum_{i=sys}\lambda_i\mathbf{\Psi}_i\, ,
\end{eqnarray}
where the vector $\mathbf{\Psi}_i$ is the template for systematic $i$.  We then project out the systematics by adding to the covariance matrix $\mathbf{C}$ in both the quad vector and Fisher matrix (see Eq.~\ref{E:fq}) the outer product of each systematic template with itself, according to
\begin{eqnarray} \label{E:syscov}
\mathbf{C}=\mathbf{C^{\rm true}}+\sum_{i=sys}\zeta_i \mathbf{\Psi}_i\mathbf{\Psi}_i^T\, ,
\end{eqnarray}
which yields an unbiased estimator when $\zeta\to\infty$.  We set $\zeta_i$ for each template large enough such that the resulting $F_{ij}$ and $q_i$ converge.

\subsection{Templates} \label{S:system}

In this section we enumerate the various systematics templates for which we search in RQCat.  The systematics we consider are extinction, stellar contamination, red stellar contamination, stellar color locus offset, stellar flux-error locus offset, air mass, seeing, sky brightness, modified Julian date, camera column, and atmospheric refraction.  We describe the possible effects of each systematic below, as well as the construction of the templates.  We list the template numerical properties, including the template mean, standard deviation, $\zeta$, and $\delta_q$ vs.~$\Psi$ slopes, in Table \ref{T:sys}.  For air mass, seeing, sky brightness, MJD, camera column, and atmospheric refraction, the required values were given per SDSS scanning field, where each field is a spherical rectangle on the sky.  We used the MANGLE \citep{Swanson:2007aj} software to convert these fields to HEALPix pixels.  Also, note that each template had its mean subtracted.

\textit{Extinction (ext):} The Galactic extinction was taken from the infrared dust maps described in \citet{Schlegel:1997yv} and given in terms of the reddening $E(B-V)$ in magnitudes.  The quasar catalog can be sensitive to any errors in the normalization or shape of the extinction law $A_\lambda/E(B-V)$.  Also, the extinction correction is sensitive to the intra-band spectral energy distributions (SED) of the source under consideration; quasars often have extreme intraband colors and this could lead to a variation of the quasar density determined from ``extinction-corrected'' photometry with $E(B-V)$.

\textit{Stellar contamination (star):}  We use the stellar overdensity map mentioned in Sec.~\ref{S:data} as the template.  When constructing a quasar catalog, the main challenge is to separate stars from quasars, since both appear as point sources.  The stellar contamination rate for RQCat is $\sim$3\% at lower redshifts, yet more precision may be required for non-Gaussianity measurements.

\textit{Red stellar contamination (rstar):} The template is similar to stellar contamination, except the stellar sample is narrowed further to include only those that also satisfy the color cut $g-r>1.4$.  Note that this template will also project our contamination from blue stars, since the ``blue star'' map is a linear combination of ``stars'' and ``red stars.''

\textit{Stellar color locus offset (ugr, gri, riz):} This is sensitive to position-dependent offsets in the photometric calibration of the survey.  We can measure these variations through shifts in the stellar locus, or scatter plots of the stars between two colors.  Templates were constructed, in terms of the ordinate vs.~abscissa of the plot, for color loci $g-r$ vs.~$u-g$ ($ugr$), $r-i$ vs.~$g-r$ ($gri$), and $i-z$ vs.~$r-i$ ($riz$) in magnitudes.  We start by determining the color locus for the sample of stars used in our stellar overdensity map, which we then fit to a line.  We then divide the sample into 49,152 HEALPix $N_{\rm side}=64$ pixels, with each large pixel containing 16 $N_{\rm side}=256$ small pixels we use in our analysis so that there are sufficient stars to measure the color locus offset.  In each large pixel, we use its stars to construct a color locus for that pixel.  We then take the median of the difference between each star's ordinate in the large pixel's color locus with the linear fit of the entire sample's color locus as a measure of the offset.  We then give the offset for all 16 small pixels the value of the offset calculated from the large pixel.

\textit{Stellar flux-error locus offset (uerr):} Statistical photometric (flux) errors can scatter objects into or out of the region of color space used for quasar selection (or the region of color space that maps to a particular photo-$z$ range), and thus affect the density of selected quasars. Variations of these errors across the sky can then produce spurious large scale power in the quasar maps. A template was used only for the $u$ band. We construct the template in a manner similar to the stellar color locus offset template, except that (1) the ordinate and abscissa for the stellar loci are the flux error and flux (in magnitudes), respectively, and (2) instead of a linear fit, we just fit the stellar locus to a horizontal line (which is the same as averaging the ordinate values).

\textit{Air mass (airu):} Air mass is the path length of atmosphere a photon travels from its source to the telescope, normalized to unity at zenith; it depends on both sky position and observation time. Increased airmass increases both the statistical errors on the magnitudes of objects, as well as producing color terms since the bluest wavelengths are selectively attenuated. The 5 filters in SDSS are taken in rapid succession, so taking the air mass value from the $u$ band observation should be sufficient for template projection purposes.  The values for the template are taken from the Catalog Archive Server (CAS) \citep{York:2000gk}.

\textit{Seeing (seeu, seer):} Seeing is defined here as the FWHM of the point spread function (PSF) in a pixel, given in arcseconds from the SDSS CAS.  It affects the photometric noise, de-blending of sources, and star-galaxy (i.e. point source-extended source) separation. Templates were constructed for the $u$ ($seeu$) and $r$ ($seer$) bands.

\textit{Sky brightness (skyu, skyi):} Sky brightness is namely the brightness of the atmosphere during observation.  Small-scale variations in the sky brightness can affect object identifications, in that the sky brightness is subtracted from the bands before identification is performed.  The values, given in maggies/arcsec$^2$, are taken from the SDSS CAS.  Templates were constructed for the $u$ band ($skyu$), in which sky brightness is dominated by moonlight if present, and the $i$ band ($skyi$), in which sky brightness is dominated by light from the intrinsic atmosphere. 

\textit{Modified Julian Date (mjd):} The modified Julian date (MJD) is a linear unit of time, given in days, to track astronomical observations.  The MJD values for the SDSS scanning fields were taken from the SDSS CAS.  Mode-projecting this template will project out modes from any systematic that varies linearly with observation time.

\textit{Camera Column (cam1...cam5):} The CCD camera for SDSS consists of six (6) camera columns (camcols).  The CCDs are highly calibrated; however, an unknown malfunction in a camcol could misidentify objects, contaminating the quasar density field.  The camcols for the SDSS scanning fields are given in the SDSS CAS.  A template for each camcol is \textit{almost} binary, in that for camcol $j$, the template value for a pixel is 1 if the pixel was observed by camcol $j$ and 0 if it was not.  The ``almost'' is because the HEALPix pixels cross the scanning field boundaries, causing cases where a pixel is observed by multiple fields.  In these cases the template value for camcol $j$ is given a fractional value denoting the percentage \textit{of the observed part} of the pixel that was observed by camcol $j$.  For example, lets say a pixel was observed by two fields with different camcols, where camcols 2 and 4 observed 60\% and 35\% of the pixel, respectively, and 5\% of the pixel was either not observed at all or observed by a bad field which had to be culled from the data.  In this case, the templates for camcols 2 and 4 would have values 60/(60+35)=0.63 and 35/(60+35)=0.37, respectively.  It is easy to see that the sum of all the camcol templates must be a vector with all ones, which we call $I$\footnote{$I$ in this paper is \textit{not} equal to the identity matrix.}.  Thus, the template for camcol 6 is linearly dependent on the templates for camcols 1--5, making it unable to remove contaminated modes that are independent of the other camcols.  Therefore we only mode-project camcols 1--5.  We also orthogonalize these templates to speed the convergence of the estimator.

\textit{Atmospheric Refraction (ref):} Refraction of photons through the atmosphere can perturb the position of a source, distorting the density field and increasing/decreasing correlations.  Since refraction is chromatic, the different bands will give different positions to a source.  The template is designed to measure the elongation of the refraction-induced ``rainbow'' in the scan direction. The template is given in terms of the altitude ($h$) and position angle ($\phi$) of the telescope as $\cot^2(h)\cos(2\phi)$, which is proportional to the quadrupole field that describes the perturbations due to refraction.  The $h$ and $\phi$ values for each scanning field were taken from the SDSS Data Archive Server (DAS).

While each of the systematic templates has a ``primary motivation,'' each one may be sensitive to several contaminating effects. For example, the stellar color locus test is obviously sensitive to photometric calibration errors, but also contains some sensitivity to Galactic dust and to the variation of stellar populations over the sky. The stellar density maps were motivated by the possibility of stellar contamination, however they are also correlated with the dust map since both are largest at low Galactic latitudes. However, each of these effects is independent of the true quasar density and correlations of the quasar map with any of them would be problematic; note also that since the template projection removes any linear combination of the input templates, there is no requirement that these templates be orthogonal, or sensitive to only a single root cause of systematic error.

We also list the Pearson correlation coefficients for all the templates in Table \ref{T:corr}.  We see that almost all the templates have low correlations with other templates.  Of the 171 possible template pairings, only 7 of them have correlations (positive or negative) greater than $\pm$20\%.  The only two templates correlated with each other by more than 50\% are the seeing templates in the $u$ and $r$ bands (80\%).  This implies that mode-projecting this set of the templates should remove a diverse set of modes.

Note that if the range of variation in conditions is large (e.g. as for the airmass or sky brightness), the spurious contribution to the quasar density may be nonlinear in the template -- e.g. we may have
\begin{equation}
n_{\rm spurious} = {\rm const} + \alpha (X-\bar X) + \beta (X-\bar X)^2 + ...,
\end{equation}
where $X$ is a template (e.g. the airmass) and $\beta$ is a nonlinearity coefficient.
Since the template projection method is a linear method, we can only remove the $\beta$-term by adding yet another template proportional to $X^2$; in general, if multiple templates have large variations we would also have to include products of template maps. We tested the mode-projection of products of templates for the cross-correlation of z01 and z02 since it has the larger cross-correlation.

\section{Results} \label{S:results}

The cross-correlation power spectra between the redshift slices before mode projection are shown in Fig.~\ref{F:qcross}. Out of 12 cross-power measurements (4 $\ell$-bins times 3 cross-correlations), two exhibit significant cross-powers.  The largest one is between slices z01 and z02, or $C_\ell^{12}$, with a value of $(1.63\pm0.18)\times10^{-4}$ ($9.2\sigma$) in the first $\ell$-bin ($2\leq \ell<7$).  We see there is also a sizable cross-correlation between slices z01 and z03, or $C_\ell^{13}$, appearing in the second $\ell$-bin ($7\leq \ell<12$) with a value of $(7.5\pm2.8)\times10^{-5}$ (2.7$\sigma$).  Large-scale biases are not expected to be this high, particularly with the magnitude of $C_\ell^{12}$.  This leads us to suspect that systematic errors in these slices are corrupting these quasar maps.  The redshift slices z02 and z03 exhibit no significant cross-correlation.

After mode-projecting each of our candidate systematic templates, we find several templates that significantly increased the cross-correlations.  We list these templates for both cross-correlations in Tables \ref{T:sysf1} and \ref{T:sysf2}.  The biggest contributors by far to $C_\ell^{12}$ seem to be seeing in the $u$ and $r$ bands.  Mode-projecting each of these individually reduce the cross-correlation significance by more than 3$\sigma$.  Red star contamination, photometric calibration errors, and air mass also seem to play a role.  Our uncovering of red star contamination seems to be in agreement with \citet{Slosar:2008hx}, which found both of their samples of quasars to be contaminated by them.  Mode-projecting all these templates yields a cross-correlation measurement of $(1.03\pm0.19)\times10^{-4}$ ($5.4\sigma$), which is closest to the value from $u$-band seeing.  It appears that seeing is a very important systematic effect in our quasar maps, and it should be monitored in future LSS experiments.  We tried to decrease $C_\ell^{12}$ further by mode-projecting products of templates.  We tested several combinations, including ($airu$)$^2$, ($ugr$)$^2$, ($star$)$^2$, ($seeu$)$^2$, ($seer$)$^2$, $ugr-riz$, $ugr-airu$, $riz-airu$, $ugr-seeu$, $riz-seeu$, $airu-seeu$, $ugr-seer$, $riz-seer$, and $airu-seer$\footnote{This list is not exhaustive of all the possible combinations, but it is large subset of the ones we would expect to be important, including the combinations between the two seeing templates.}  None of the combinations we tested decreased the cross-correlation significantly.  It is important to note that although we were able to reduce the cross-correlation significance by 3.8$\sigma$, the cross-correlation is far too high.  We conclude that more systematics are present that were not found.

For $C_\ell^{13}$, stellar contamination is the dominant systematic.  Mode-projecting this template reduces the cross-correlation significance by more than half a $\sigma$, enough to drive the significance below 2$\sigma$.  Methods to reduce stellar contamination should also be developed for future LSS surveys.  Seeing systematics seem to be having a small affect on this cross-correlation measurement as well.  When we mode-project stellar contamination and seeing, we find a cross-correlation value of $(5.0\pm2.8)\times10^{-5}$ ($1.8\sigma$).  Final plots for $C_\ell^{12}$ and $C_\ell^{13}$ are shown in Fig.~\ref{F:clqcrsys}.

Note we can use the measured cross-correlations to estimate the level of contamination in the redshift slices.  For example, we can estimate the z01-z02 cross-correlation due to contaminants as $\ell(\ell+1)C_\ell^{12}/(2\pi)\sim\VEV{(\delta n/n)^2}\sim f_{\rm cont}^2$, where $f_{\rm cont}$ is the fraction of contaminants. This implies that the contamination levels before mode projection for $C_\ell^{12}$, $C_\ell^{13}$, and $C_\ell^{23}$ in the relevant contaminated $\ell$-bins\footnote{For $C_\ell^{23}$ I simply use the first $\ell$ bin.} are (2.2$\pm$0.7)\%, (3.3$\pm$2.0)\%, and $<$2.5\%, respectively.  After mode projection, the contamination levels for $C_\ell^{12}$ and $C_\ell^{13}$ reduce to (1.8$\pm$0.8)\% and (2.7$\pm$2.0)\%, respectively.  From the theoretical auto-correlation plots in Fig.~\ref{F:cl}, it appears we need the contamination level to be $\sim$1\% or less to see fluctuations from $\fnl\sim100$ or 0.6\% for $\fnl\sim10$.

It is important to note that we can divide the systematic errors in our photometric quasar maps into two categories -- {\em contamination} and {\em calibration}. Stellar contamination is the principal example of the former since we are selecting point sources: it results due to the imperfection of the photometric redshift method in identifying quasars. (In principle, crowding could lead to ``negative'' contamination since the presence of a higher stellar density may reduce the completeness of the catalog.)  \citet{Richards:2008eq} claim a 96\% efficiency in identifying UVX quasars with $z<2.2$, yet this probably decreases for higher redshift quasars\footnote{The efficiency decreases to 46\% for all higher redshift quasars, but it is probably much higher than that for UVX quasars.  \citet{Richards:2008eq} does not state the efficiency in this case.}.  This may explain why stellar contamination arises in the cross-correlation of z01--z03 and not z01--z02.

The other type of systematic error is calibration, where variations in the observing conditions or instrument properties propagate to changes in the completeness of the quasar catalog. This may consist of first-order calibration effects (as in the case of photometric calibration errors, or airmass effects that shift the center of the $u$ band), as well as second-order or noise-induced effects such as that contributed by the sky brightness.
Although contamination has been the main focus of previous photometric catalogs, it appears that even the excellent internal calibration of the SDSS does not provide samples with the uniformity on large scales needed for $\fnl$ studies.

\section{Conclusions} \label{S:conclude}

We searched for various systematic effects in the SDSS DR6 photometric quasar catalog \citep{Richards:2008eq}, which is the latest catalog available.  We find that the clustering signal from this catalog is contaminated on the largest scales.  In particular, our z01 subsample ($0.9<z<1.3$) and our z02 subsample ($1.6<z<2.0$) exhibits a 9.2$\sigma$ detection of cross correlation on scales $2\leq\ell<7$, even though quasars from separate redshifts on these scales should be uncorrelated.  We suspect that the cause of this large cross-correlation is systematic effects.  By mode-projecting several candidate templates individually, we find that the maps are indeed contaminated by stars, photometric calibration errors, and variations in air mass and seeing, with seeing being the dominant contaminant.  By mode-projecting this group of templates together, we were able to decrease the significance of the cross-correlation by more than 3$\sigma$.  However, it is still too high to provide a competitive constraint on $f_{\rm NL}$
through scale-dependent bias. We find spurious number density fluctuations of $\sim 2$\% rms, whereas we need a contamination level less than 1\% (0.6\%) in order to measure values of $\fnl$ less than 100 (10).  We recommend new efforts into improving photometry in order to use photometric quasars to constrain nongaussianity on large scales.

\acknowledgments
We thank G.~Richards, S.~Ho and O.~Dor\'{e} for useful discussions and comments. AP was supported during part of this research by DOE DE-FG03-92-ER40701, NASA NNG05GF69G, the Gordon and Betty Moore Foundation, and a NASA Einstein Probe mission study grant, ``The Experimental Probe of Inflationary Cosmology.''  AP is currently supported by an appointment to the NASA Postdoctoral Program at the Jet Propulsion Laboratory, California Institute of Technology, administered by Oak Ridge Associated Universities through a contract with NASA.  CH is supported by the US Department of Energy under contract DE-FG03-02-ER40701, the Alfred P Sloan Foundation, and the David and Lucile Packard Foundation.

Funding for the SDSS and SDSS-II has been provided by the Alfred P. Sloan Foundation, the Participating Institutions, the National Science Foundation, the U.S. Department of Energy, the National Aeronautics and Space Administration, the Japanese Monbukagakusho, the Max Planck Society, and the Higher Education Funding Council for England. The SDSS Web Site is {\tt http://www.sdss.org/}. The SDSS is managed by the Astrophysical Research Consortium for the Participating Institutions. The Participating Institutions are the American Museum of Natural History, Astrophysical Institute Potsdam, University of Basel, University of Cambridge, Case Western Reserve University, University of Chicago, Drexel University, Fermilab, the Institute for Advanced Study, the Japan Participation Group, Johns Hopkins University, the Joint Institute for Nuclear Astrophysics, the Kavli Institute for Particle Astrophysics and Cosmology, the Korean Scientist Group, the Chinese Academy of Sciences (LAMOST), Los Alamos National Laboratory, the Max-Planck-Institute for Astronomy (MPIA), the Max-Planck-Institute for Astrophysics (MPA), New Mexico State University, Ohio State University, University of Pittsburgh, University of Portsmouth, Princeton University, the United States Naval Observatory, and the University of Washington.

\copyright 2013.  All rights reserved.

\appendix

\section{Quasar redshift distributions} \label{A:redproc}

In \citet{Ho:2008bz}, photometric redshift distributions $f_i(z)$ for quasars were constructed using spectroscopic data from 2SLAQ \citep{Richards:2005qh}.  These redshift distributions determine how the matter overdensity $\delta(\vecx)$ relates to the quasar overdensity $\delta_q(\hatn)$
\begin{eqnarray} \label{E:matterqso}
\delta_q(\hatn) = \int_0^\infty f(z)\delta[\hatn,\chi(z)]dz\, .
\end{eqnarray}
We refer the reader to \citet{Ho:2008bz} for the theory behind this method, and we describe our method which is similar to and follows from \citet{Ho:2008bz}.

The expression for $f(z)$ is given as
\begin{eqnarray} \label{E:reddistbasic}
f_i(z)=b(z)\Pi(z)+\int_z^\infty W(z,z')[\alpha(z')-1]\Pi_i(z')dz'\, ,
\end{eqnarray}
where $b(z)$ is the linear bias as a function of redshift, $\chi(z)=\int_0^z c/H(z')dz'$ is the comoving 
radial distance, and $\Pi_i(z)$ is the probability distribution for the quasar redshifts.  The second term in Eq.~\ref{E:reddistbasic} is due to magnification bias, which becomes important for large redshifts, with the lensing window function $W(z,z')$ given for a flat universe by
\begin{eqnarray} \label{E:wzz}
W(z,z') = \frac{3}{2}\Omega_m H_0^2\frac{1+z}{cH(z)}\chi^2(z)\left[\frac{1}{\chi(z)}-\frac{1}{\chi(z')}\right]\, ,
\end{eqnarray}
and $\alpha(z)$ being the logarithmic slope of the number counts of quasars as a function of flux: $N(>F)\propto F^{-\alpha}$.  In this appendix, we describe the method of calculating the various factors in Eq.~\ref{E:reddistbasic} that comprise $f(z)$.

We rely on spectroscopic data to compose redshift distributions.  Specifically, we use spectroscopic quasars (spectro-QSOs) from an area with high spectroscopic coverage to construct a preliminary probability distribution $\Pi_{i,\rm prelim}$ and $\alpha(z)$.  We use spectroscopic data from the 2SLAQ survey, which contains 8389 spectro-QSOs over its total region of view.  We restrict ourselves to using spectroscopic data from five rectangles on the sky with declination range $-01^\circ00'36''-00^\circ35'24''$ and right ascension ranges $137^\circ-143^\circ$, $150^\circ-168^\circ$, $185^\circ-193^\circ$, $197^\circ-214^\circ$, and $218^\circ-230^\circ$, the same as those used in \citet{Ho:2008bz}, for the quasar analysis.  These rectangles in particular have high spectroscopic coverage and contain 5383 quasars.  Since we also need photo-$z$s for the quasars to construct probability densities, we only use quasars that have matches in the RQCat, decreasing the number of objects to 3443.

\begin{figure}
\begin{center}
{\scalebox{.5}{\includegraphics{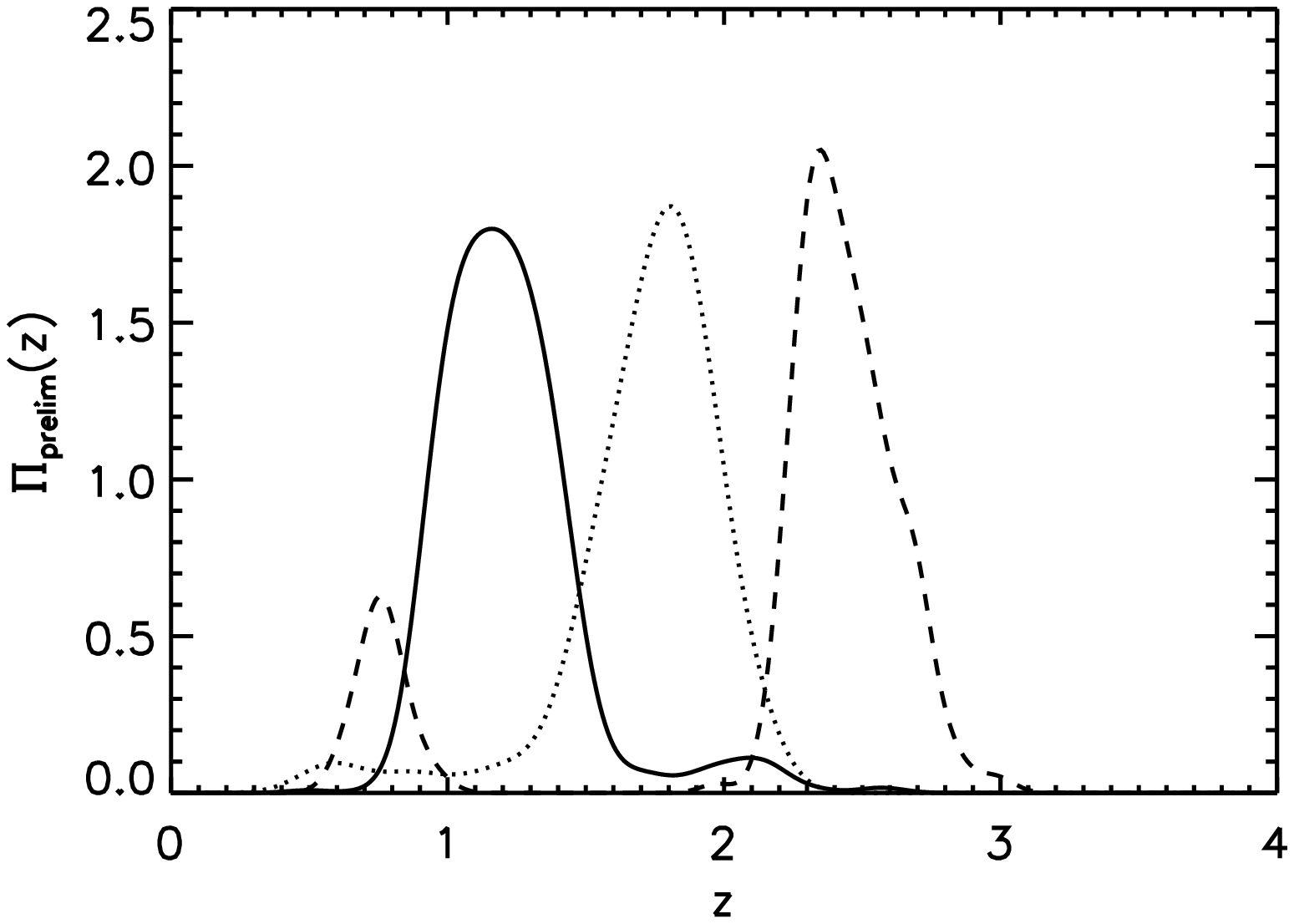}}}
\caption[The preliminary redshift distributions for the QSO photometric redshift slices]{The preliminary redshift distributions for the QSO photometric redshift slices z01 (solid), z02 (dotted) and z03 (dashed).  \label{F:prelim}}
\end{center}
\end{figure}

We calculate $\Pi_{i,\rm prelim}(z)$ for each photometric redshift slice using a kernel density estimator of the form
\begin{eqnarray} \label{E:prelim}
\Pi_{i,\rm prelim}(z) = \frac{1}{N_q}\sum_{k=1}^{N_q}\frac{1}{\sqrt{2\pi}\sigma}e^{-(z-z_k)^2/2\sigma^2}\, ,
\end{eqnarray}
where $N_q$ is the number of spectro-QSOs (matched with a photo-QSO in RQCat) in the photometric redshift slice, $z_k$ is the spectro-$z$ of the $k$th-matched quasar, and $\sigma$ is the slice's kernel width.  $\sigma$ is chosen to be smaller than any real features in $\Pi_{i,\rm prelim}$ yet large enough to smooth out shot noise.  Table~\ref{T:prelimstat} lists $N_q$ and $\sigma$ for each redshift slice, and Fig.~\ref{F:prelim} shows a plot of $\Pi_{i,\rm prelim}$ for each slice.  We also calculate $\alpha$, the logarithmic slope of the number counts of quasars in terms of flux, by creating a histogram of number counts in terms of the PSF magnitude in the $g$-band around $g=21$ and calculating the \emph{actual} slope around this value.  This value for each redshift slice is also listed in Table \ref{T:prelimstat}.

The expression for $f_i(z)$ in Eq.~\ref{E:reddistbasic} requires the true probability distribution $\Pi_i(z)$.  However, since $\alpha-1$ is small, the second term is subdominant to the first and we can substitute $\Pi_i(z)$ with $\Pi_{i,\rm prelim}(z)$ in the second term, giving us
\begin{eqnarray} \label{E:reddistapprox}
f_i(z)\simeq b(z)\Pi_i(z)+\int_z^\infty W(z,z')[\alpha(z')-1]\Pi_{i,\rm prelim}(z')dz'\, .
\end{eqnarray}
$\Pi_i(z)$ cannot be similarly substituted for in the first term, so we must estimate $b(z)\Pi_i(z)$ using LSS data.  We estimate $b(z)D(z)$ as nearly constant, writing
\begin{eqnarray} \label{E:amp}
b(z)\Pi_i(z)D(z)=A_i\Pi_{i,\rm prelim}\, ,
\end{eqnarray}
where $A_i$ is a constant.  Generally, $A_i$ would be a piecewise function of $z$ with several breaks in order to estimate $A_i$ precisely; however, we were able to estimate $A_i$ as a constant function with very small uncertainties.

\begin{figure}
\begin{center}
{\scalebox{.5}{\includegraphics{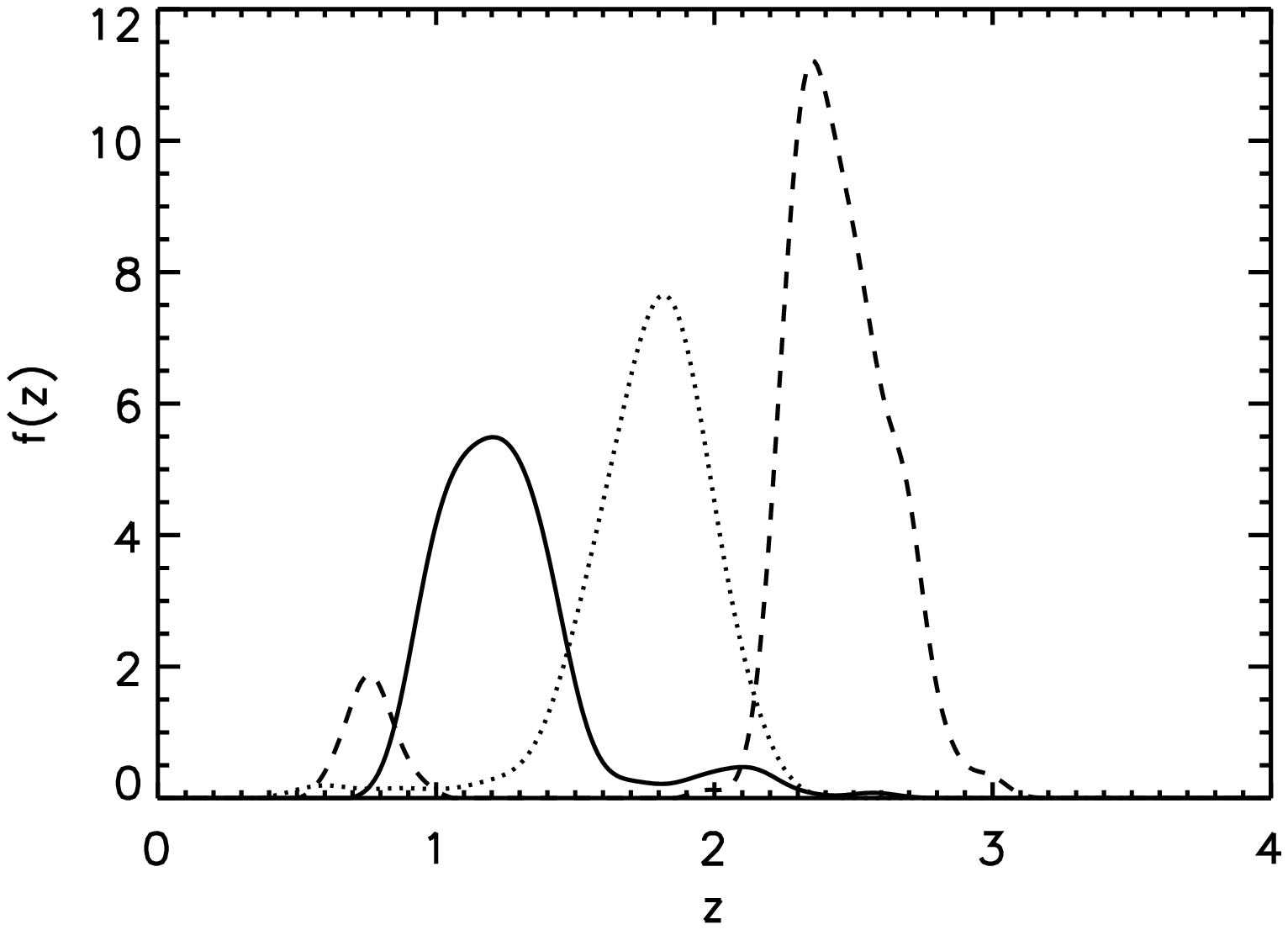}}}
\caption[The redshift distributions for the QSO photometric redshift slices]{The redshift distributions for the QSO photometric redshift slices z01 (solid), z02 (dotted) and z03 (dashed).  \label{F:redplot1}}
\end{center}
\end{figure}

We estimate $A_i$ in each redshift slice by constraining its effect on the quasar clustering.  We begin by estimating the quasar correlation function $w_i(\theta)$ in each redshift slice using the method presented in \citet{Landy:1993yu}.  Specifically, we use the estimator $\hat{w}_4(\theta)$ along with its variance given in \citet{Landy:1993yu}, given by
\begin{eqnarray} \label{E:wthest}
\hat{w}_4(\theta)=\frac{DD(\theta)-2DR(\theta)+RR(\theta)}{RR(\theta)}\, ,
\end{eqnarray}
where $DD(\theta)$, $DR(\theta)$, and $RR(\theta)$ are properly normalized histograms of the number of data-data pairs, data-random pairs, and random-random pairs, respectively, binned in terms of angular separation.  We use the following expression for the Poisson uncertainty in our estimator
\begin{eqnarray} \label{E:wthesterr}
{\rm var}[\hat{w}_4] = \frac{2}{n_r(n_r-1)RR}+\frac{2}{n(n-1)DD}\, ,
\end{eqnarray}
where $n$ ($n_r$) is the number of data (random) points used to calculate $DD(\theta)$ ($RR(\theta)$).  Note that for the stochastic calculations we use 25 simulations with the number of random points equal to twice the number of data points.  We calculate $w_i(\theta)$ in 10 logarithmic bins of equal (logarithmic) size in the range $0.3^\circ < \theta < 6^\circ$.  We avoid estimating $w_i(\theta)$ at smaller angles due to nonlinearities and at larger angles due to possible non-gaussianity effects on large scales.  We compare this estimate of $w_i(\theta)$ to the expression 
\begin{eqnarray}\label{E:wth}
w(\theta)&=&\int_0^\infty dk\,k P(k) F(k,\theta)\nonumber\\
F(k,\theta)&=&k\sum_{l=1}^\infty\left(\frac{2l+1}{2\pi^2}\right)P_l(\cos\theta)[W_l(k)]^2\, ,
\end{eqnarray}
where we substitute Eq.~\ref{E:amp} for the first term in Eq.~\ref{E:reddistapprox} to calculate $W_l(k)$ in terms of $A$.  We include redshift-space distortions in $W_l(k)$.  For calculating $F(k,\theta)$ we use the Limber approximation for $k>0.0155$ Mpc$^{-1}$\footnote{We verified that for $k>0.017$ Mpc$^{-1}$ the summand vanishes for $l<40$, which removes all terms over which the Limber approximation is not valid.} by converting the sum to an integral, replacing $P_l(\cos\theta)\to J_0(l\theta)$, and replacing $W_l(k)\to \sqrt{\pi/2l}f(l/k)/k$.
We can write $w_i(\theta)$ as a sum of terms linear and quadratic in $A_i$, which allows us to fit for this parameter for each redshift slice.  We also add to this model a constant term $B_i$ due to systematics effects for which we also fit\footnote{Note that we did not include $B_i$ for redshift slice z03 because it caused $A_i$ to not be constrainable.}.  In Table \ref{T:ai}, we list estimates for the $A_i$, from which we derive the redshift distributions $f_i(z)$ shown in Fig.~\ref{F:redplot1}.

\begin{table}
\caption[Properties of the 3 QSO photometric redshift slices]{\label{T:redshifts1} Properties of the 3 QSO photometric redshift slices; $z_{\rm p}$ is the photometric redshift range, and $z_{\rm mean}$ is the mean (true) redshift of the slice, and $N_{qso}$ is the number of QSOs in the redshift slice.}
\begin{center}
\item[]\begin{tabular}{@{}ccccc}
\toprule \toprule Label&$z_{\rm p}$&$z_{\rm mean}$&$N_{qso}$\\
                                  \midrule z01&0.9-1.3&1.230&75,835\\
                                  z02&1.6-2.0&1.731&91,356\\
                                  z03&2.3-2.9&2.210&10,806\\ \bottomrule
\end{tabular}\end{center}
\end{table}

\begin{table}
\caption{\label{T:sys} Template properties for various systematics.  We list each template's mean ($\overline{\Psi}$), standard deviation ($\sigma_\Psi$), prefactor ($\zeta$) for mode-projecting each template, and slope with each of quasar distributions from the three redshift slices.  The slope (with appropriate units) of a template versus quasar distribution $i$ is equal to the cross-correlation coefficient between them times $\sigma_\delta/\sigma_\Psi$, where $\sigma_\delta\sim1/\sqrt{n}$.  Note that the mean is subtracted from each template before the analysis commences.}
\begin{center}
\texttt{\tiny
\item[]\begin{tabular}{p{26mm}p{18mm}p{14mm}p{5mm}p{24mm}p{26mm}p{24mm}}
\toprule \toprule Systematic (units)&$\overline{\Psi}$&$\sigma_\Psi$&$\zeta$&z01 slope&z02 slope&z03 slope\\
                                  \midrule ebv (mag)&0.0251&0.0109&10$^7$&+0.415$\pm$0.424&-0.249$\pm$0.351&-5.84$\pm$2.98\\
                                  star&$+2.96\times10^{-7}$&0.658&10$^6$&+0.0149$\pm$0.0070&+0.0121$\pm$0.0058&+0.130$\pm$0.049\\
                                  rstar&$+3.67\times10^{-6}$&1.38&10$^6$&(+$2.60\pm3.35)\times10^{-3}$&(+$7.29\pm2.77)\times10^{-3}$&(+$3.89\pm23.58)\times10^{-3}$\\
                                  ugr (mag)&+0.0176&0.0354&10$^9$&-0.0453$\pm$0.1306&-0.276$\pm$0.108&+0.438$\pm$0.919\\
                                  gri (mag)&+$6.59\times10^{-3}$&0.0282&10$^9$&-0.0852$\pm$0.1639&(-$8.98\pm135.61)\times10^{-3}$&-0.322$\pm$1.154\\
                                  riz (mag)&+$1.84\times10^{-3}$&0.0211&10$^7$&+0.201$\pm$0.219&+0.401$\pm$0.181&-0.0270$\pm$1.5419\\
                                  uerr (mag)&-0.0164&0.0207&10$^5$&-0.480$\pm$0.223&-0.0589$\pm$0.1847&-2.54$\pm$1.57\\
                                  airu (mag)&1.15&0.0921&10$^6$&-0.0583$\pm$0.0502&-0.0700$\pm$0.0415&-0.0838$\pm$0.3532\\
                                  seeu (arcsec)&1.54&0.175&10$^6$&-0.176$\pm$0.026&-0.124$\pm$0.022&-0.198$\pm$0.186\\
                                  seer (arcsec)&1.35&0.164&10$^6$&-0.162$\pm$0.028&-0.112$\pm$0.023&-0.130$\pm$0.198\\
                                  skyu (maggies/arcsec$^2$)&22.2&0.230&10$^6$&(-$3.25\pm20.10)\times10^{-3}$&+0.0265$\pm$0.0166&-0.170$\pm$0.141\\
                                  skyi (maggies/arcsec$^2$)&20.3&0.216&10$^6$&(-$9.28\pm21.40)\times10^{-3}$&+0.0464$\pm$0.0177&-0.0936$\pm$0.1506\\
                                  mjd (days)&52505&552&10$^2$&(+$3.4\pm8.4)\times10^{-6}$&(+$1.92\pm0.69)\times10^{-5}$&(+$1.52\pm0.59)\times10^{-4}$\\
                                  cam1&$+1.19\times10^{-9}$&$2.85\times10^{-3}$&10$^6$&-2.08$\pm$1.62&+1.45$\pm$1.34&+1.33$\pm$11.42\\
                                  cam2&$-7.11\times10^{-10}$&$2.85\times10^{-3}$&10$^6$&-0.324$\pm$1.622&-0.697$\pm$1.342&+0.125$\pm$11.416\\
                                  cam3&$-8.42\times10^{-10}$&$2.85\times10^{-3}$&10$^6$&+0.977$\pm$1.622&+1.72$\pm$1.34&+7.87$\pm$11.42\\
                                  cam4&$-1.46\times10^{-9}$&$2.85\times10^{-3}$&10$^6$&+4.44$\pm$1.62&-4.73$\pm$1.34&+8.80$\pm$11.42\\
                                  cam5&$-2.92\times10^{-9}$&$2.85\times10^{-3}$&10$^6$&-3.32$\pm$1.62&-1.97$\pm$1.34&-7.72$\pm$11.42\\
                                  ref&-0.0596&0.439&10$^6$&(-$2.59\pm10.53)\times10^{-4}$&(+$5.00\pm8.71)\times10^{-3}$&+0.0299$\pm$7.41\\
                                  \bottomrule
\end{tabular}}\end{center}
\end{table}

\begin{table}
\caption{Correlation coefficients between templates.  Template abbreviations given in Sec.~\ref{S:system}.\label{T:corr}}
\texttt{\tiny
\begin{tabular}{
p{4.1mm}|p{4.1mm}p{4.1mm}p{4.1mm}p{4.1mm}p{4.1mm}p{4.1mm}p{4.1mm}p{4.1mm}p{4.1mm}p{4.1mm}p{4.1mm}p{4.1mm}p{4.1mm}p{4.1mm}p{4.1mm}p{4.1mm}p{4.1mm}p{4.1mm}p{4.1mm}
}
\hline\hline
 & ebv & star & rstar & ugr & gri & riz & uerr & airu & seeu & seer & skyu & skyi & mjd & cam1 & cam2 & cam3 & cam4 & cam5 & ref \\
\hline
ebv&+1.000&+0.327&+0.118&+0.016&-0.068&-0.252&-0.140&+0.004&-0.043&-0.087&-0.025&-0.069&+0.002&+0.017&-0.004&-0.010&+0.006&-0.008&-0.092\\
star&+0.327&+1.000&+0.364&+0.254&+0.082&-0.128&-0.182&-0.036&-0.164&-0.140&-0.011&-0.086&-0.006&-0.013&+0.024&+0.026&+0.009&-0.014&+0.011\\
rstar&+0.118&+0.364&+1.000&+0.082&+0.065&+0.035&+0.132&-0.056&-0.177&-0.127&+0.050&+0.005&+0.030&-0.026&+0.015&+0.005&+0.001&-0.022&+0.006\\
ugr&+0.016&+0.254&+0.082&+1.000&+0.006&+0.018&+0.020&-0.006&+0.013&+0.011&-0.002&-0.016&-0.039&-0.015&+0.065&+0.019&+0.021&-0.009&+0.052\\
gri&-0.068&+0.082&+0.065&+0.006&+1.000&-0.086&+0.121&-0.019&+0.024&+0.024&+0.079&+0.073&-0.012&-0.010&-0.040&-0.003&-0.003&+0.004&+0.035\\
riz&-0.252&-0.128&+0.035&+0.018&-0.086&+1.000&+0.133&-0.103&-0.110&-0.021&+0.137&+0.084&+0.013&+0.002&+0.067&-0.018&-0.050&-0.019&+0.039\\
uerr&-0.140&-0.182&+0.132&+0.020&+0.121&+0.133&+1.000&-0.037&+0.012&+0.031&+0.116&+0.102&+0.016&-0.018&-0.003&+0.016&+0.019&-0.020&+0.031\\
airu&+0.004&-0.036&-0.056&-0.006&-0.019&-0.103&-0.037&+1.000&+0.135&+0.114&-0.099&-0.041&-0.006&-0.016&+0.003&-0.001&-0.016&-0.009&-0.076\\
seeu&-0.043&-0.164&-0.177&+0.013&+0.024&-0.110&+0.012&+0.135&+1.000&+0.804&-0.119&-0.040&-0.123&+0.033&-0.065&-0.012&-0.020&+0.010&-0.027\\
seer&-0.087&-0.140&-0.127&+0.011&+0.024&-0.021&+0.031&+0.114&+0.804&+1.000&-0.127&-0.058&-0.149&+0.040&-0.022&-0.002&-0.019&-0.017&-0.013\\
skyu&-0.025&-0.011&+0.050&-0.002&+0.079&+0.137&+0.116&-0.099&-0.119&-0.127&+1.000&+0.492&+0.041&+0.008&-0.039&+0.000&+0.014&-0.026&+0.077\\
skyi&-0.069&-0.086&+0.005&-0.016&+0.073&+0.084&+0.102&-0.041&-0.040&-0.058&+0.492&+1.000&-0.151&-0.008&+0.010&+0.014&+0.031&-0.002&+0.069\\
mjd&+0.002&-0.006&+0.030&-0.039&-0.012&+0.013&+0.016&-0.006&-0.123&-0.149&+0.041&-0.151&+1.000&+0.019&-0.025&-0.041&-0.007&+0.042&-0.111\\
cam1&+0.017&-0.013&-0.026&-0.015&-0.010&+0.002&-0.018&-0.016&+0.033&+0.040&+0.008&-0.008&+0.019&+1.000&+0.000&+0.000&+0.000&+0.000&+0.026\\
cam2&-0.004&+0.024&+0.015&+0.065&-0.040&+0.067&-0.003&+0.003&-0.065&-0.022&-0.039&+0.010&-0.025&+0.000&+1.000&+0.000&+0.000&+0.000&-0.006\\
cam3&-0.010&+0.026&+0.005&+0.019&-0.003&-0.018&+0.016&-0.001&-0.012&-0.002&+0.000&+0.014&-0.041&+0.000&+0.000&+1.000&+0.000&+0.000&+0.015\\
cam4&+0.006&+0.009&+0.001&+0.021&-0.003&-0.050&+0.019&-0.016&-0.020&-0.019&+0.014&+0.031&-0.007&+0.000&+0.000&+0.000&+1.000&+0.000&+0.022\\
cam5&-0.008&-0.014&-0.022&-0.009&+0.004&-0.019&-0.020&-0.009&+0.010&-0.017&-0.026&-0.002&+0.042&+0.000&+0.000&+0.000&+0.000&+1.000&-0.001\\
ref&-0.092&+0.011&+0.006&+0.052&+0.035&+0.039&+0.031&-0.076&-0.027&-0.013&+0.077&+0.069&-0.111&+0.026&+0.006&+0.015&-0.022&+0.001&+1.000\\
\hline\hline
\end{tabular}
}
\end{table}

\begin{table}
\caption{\label{T:sysf1} Systematic templates that significantly reduced upon mode projection the cross-correlation $C_\ell^{12}$ in the first $\ell$-bin ($2\leq \ell<7$), along with the new cross-correlation value (with 1$\sigma$ errors) and the reduction in the signal-to-noise ratio (SNR).}
\begin{center}
\item[]\begin{tabular}{@{}ccc}
\toprule \toprule Systematic&$10^4C_\ell^{12}$&$\Delta$SNR ($\sigma$)\\
                                  \midrule Red stellar contamination&$1.59\pm0.19$&-0.8\\
                                  $(g-r)$ vs.~$(u-g)$&$1.60\pm0.19$&-0.8\\
                                  $(i-z)$ vs.~$(r-i)$&$1.57\pm0.19$&-0.9\\
                                  Air mass ($u$ band)&$1.66\pm0.19$&-0.5\\
                                  Seeing ($u$ band)&$1.06\pm0.19$&-3.6\\
                                  Seeing ($r$ band)&$1.16\pm0.19$&-3.1\\
                                  \bottomrule
\end{tabular}\end{center}
\end{table}

\begin{table}
\caption{\label{T:sysf2} Systematic templates that significantly reduced upon mode projection the cross-correlation $C_\ell^{13}$ in the second $\ell$-bin ($7\leq \ell<12$), along with the new cross-correlation value (with 1$\sigma$ errors) and the reduction in the signal-to-noise ratio (SNR).}
\begin{center}
\item[]\begin{tabular}{@{}ccc}
\toprule \toprule Systematic&$10^5C_\ell^{12}$&$\Delta$SNR ($\sigma$)\\
                                  \midrule Stellar contamination&$5.5\pm2.8$&-0.7\\
                                  Seeing ($u$ band)&$6.6\pm2.8$&-0.3\\
                                  Seeing ($r$ band)&$6.7\pm2.8$&-0.3\\
                                  \bottomrule
\end{tabular}\end{center}
\end{table}

\begin{table}
\caption[Properties of $\Pi_{i,\rm prelim}$ for the 3 QSO redshift slices]{\label{T:prelimstat} Properties of $\Pi_{i,\rm prelim}$ for the 3 QSO redshift slices; $N_q$ is the number of spectro-QSOs (matched with a photo-QSO in RQCat) in the photometric redshift slice, $\sigma$ is the slice's kernel width, and $\alpha$ is the logarithmic slope of number counts as a function of flux.}
\begin{center}
\item[]\begin{tabular}{@{}ccccc}
\toprule \toprule Label&$N_q$&$\sigma$&$\alpha$\\
                                  \midrule z01&599&0.08&0.48\\
                                  z02&779&0.10&0.63\\
                                  z03&236&0.06&0.64\\ \bottomrule
\end{tabular}\end{center}
\end{table}

\begin{table}
\caption{\label{T:ai} Estimates for QSO redshift distribution amplitude $A_i$ for each redshift slice with 1$\sigma$ errors.}
\begin{center}
\item[]\begin{tabular}{@{}cc}
\toprule \toprule Label&$A_i$\\
                                  \midrule z01&1.77$\pm$0.165\\
                                  z02&1.88$\pm$0.145\\
                                  z03&2.13$\pm$0.869\\ \bottomrule
\end{tabular}\end{center}
\end{table}

\end{document}